\documentclass[11pt]{article}
\usepackage{blois,epsfig}

\def\be{\begin{equation}}
\def\ee{\end{equation}}
\def\bea{\begin{eqnarray}}
\def\eea{\end{eqnarray}}

\begin{document}
\vspace*{4cm}
\title{
 INHOMOGENEOUS CHAPLYGIN GAS COSMOLOGY
 }
\author{NEVEN BILI\'{C}$^{a,b}$,
ROBERT J.\ LINDEBAUM$^c$,
GARY B.\ TUPPER$^a$, and RAOUL D.\ VIOLLIER$^a$
}
\address{$^a$Institute of Theoretical Physics and Astrophysics,
 Department of Physics, University of Cape Town,
 Private Bag, Rondebosch 7701, South Africa \\
Email: viollier@physci.uct.ac.za \\
$^b$Rudjer Bo\v skovi\'c Institute,
P.O. Box 180, 10002 Zagreb, Croatia \\
Email: bilic@thphys.irb.hr  \\
$^c$School of Chemical and Physical Sciences,
University of Natal,
Private Bag X01,
Scottsville 3209, South Africa \\
Email: lindebaumr@nu.ac.za}

\maketitle\abstracts{
The hypothesis that dark matter and dark energy are
unified through the Chaplygin gas
is reexamined.
Using a generalization of the spherical model which incorporates
effects of the acoustic horizon we show that an initially
perturbative Chaplygin gas evolves into a mixed system
containing cold dark matter in the form of gravitational condensate.
Furthermore, by including both condensate and
residual gas,
we demonstrate that the observed CMB angular and baryonic power spectra
are reproduced}

An appealing scenario in which
dark matter and dark energy are different
manifestations of a common structure ,
may be realized through the Chaplygin gas,
an exotic fluid obeying
\begin{equation}
p =  - A/\rho \; ,
\label{eq001}
\end{equation}
which has been extensively studied for its
mathematical properties \cite{jack8}.
The cosmological potential of Eq.\ (\ref{eq001}) was first
noted by Kamenshchik {\it et al} \cite{kam9} who observed that
 the Chaplygin gas interpolates between matter with
 $\rho \sim a^{-3}$
at high redshift and a cosmological constant like
$\rho \sim \sqrt{A}$ as $a$ tends to infinity.
Of particular interest is that
Eq.\ (\ref{eq001}) may be obtained from \cite{jack8,bil6,bil7}
\begin{equation}
{\cal{L}}_{\rm BI}  =
 - \sqrt{A} \; \sqrt{1 - g^{\mu \nu}
 \, \theta,_{\mu} \, \theta,_{\nu} } \; ,
\label{eq003}
\end{equation}
by evaluating the stress-energy tensor $T_{\mu \nu}$, introducing
$u_{\mu} = \theta,_{\mu} / \sqrt{g^{\alpha \beta} \,
\theta,_{\alpha} \, \theta,_{\beta} }$
for the four-velocity, and
$\rho = \sqrt{A} / \sqrt{1 - g^{\mu \nu} \,
\theta,_{\mu} \, \theta,_{\nu} }$
for the energy density. One recognizes ${\cal{L}}_{\rm BI}$
 a  Born-Infeld type Lagrangian familiar
in the $D$-brane constructions of string/$M$ theory \cite{jack8}.
 The Lagrangian (\ref{eq003}) is a special
 case of the string-theory inspired
 tachyon Lagrangian \cite{sen} in which
the constant $\sqrt{A}$ is replaced by a potential $V(\theta)$.

To be able to claim that the Chaplygin gas (or any other candidate)
actually achieves unification, one must be assured that
initial perturbations can evolve into a deeply nonlinear regime
to form a gravitational condensate of superparticles that
can play the role of cold dark matter (CDM).
In comoving coordinates,
the solution for inhomogeneous Chaplygin gas cosmology
is\cite{bil6,bil7}
\begin{equation}
\rho = \sqrt{A + B/\gamma}.
\label{eq13}
\end{equation}
Here $\gamma$ is the determinant of the induced metric
$\gamma_{ij} = g_{i0}\, g_{j0}/g_{00}-g_{ij}$,
and $B$ can be taken as constant on the
scales of interest.
Eq. (\ref{eq13}) allows us to implement the
Zel'dovich approximation~\cite{zel13}:
the transformation from Euler to Lagrange (comoving)
coordinates induces
$\gamma_{ij} = \delta_{kl}  {D_{i}}^{k}  {D_{j}}^l$,
where
${D_{i}}^{j} = a({\delta_{i}}^{j}-b{\varphi_{,i}}^{j})$
is the deformation tensor,
$\varphi$ is the velocity potential, and  the
quantity $b=b(t)$ describes the evolution of the perturbation.
The Zel'dovich approximation offers a means
of extrapolation into the
nonperturbative regime with the help of Eq.\ (\ref{eq13}) and
\begin{equation}
\gamma = a^{6}
(1-\lambda_{1}b)^2 (1-\lambda_{2}b)^2 (1-\lambda_{3}b)^2,
\label{eq17}
\end{equation}
where the $\lambda_{i}$ are the eigenvalues of
${\varphi_{,i}}^{j}$.
When one
(or more) of the $\lambda$'s is (are) positive,
a caustic forms on which $\gamma
\rightarrow 0$ and $p/\rho \rightarrow 0$, i.e.,
at the locations where structure
forms the Chaplygin gas behaves as dark matter.
Conversely, when all of the
$\lambda$'s are negative, a void forms,
$\rho$ is driven to its limiting value
$\sqrt{A}$, and the Chaplygin gas
behaves as dark energy, driving accelerated
expansion.
For the issue at hand, the Zel'dovich approximation has the
shortcoming that the effects of finite sound speed are neglected.
Indeed, in the Newtonian limit $p \ll \rho$,
an explicit solution for the perturbative
density contrast of the pure Chaplygin gas
\begin{equation}
\delta_{\rm pert} (k, a) \; \propto \;
a^{-1/4} J_{5/14}  (d_{\rm{s}} k) \, ,
\label{eq004}
\end{equation}
has been obtained \cite{fab14}.
Here $J_{\nu} (z)$ is the Bessel function, $k$ the comoving wave
number,
and $d_{\rm s}$ the comoving sonic horizon  given by
\begin{equation}
d_{\rm{s}}  = \frac{2}{7}
\frac{\left( 1 - \Omega^{2} \right)^{1/2}}{\Omega^{3/2}}
\frac{a^{7/2}}{H_0}\, ,
\label{eq005}
\end{equation}
with the equivalent matter fraction
$\Omega = \sqrt{B/(A+B)} =
\sqrt{B}/\rho_{\rm cr}$.
Thus, for $ d_{\rm{s}}k \ll 1$, $\delta_{\rm pert} \sim a$,
but for $ d_{\rm{s}}k \gg 1$,
$\delta_{\rm pert}$
undergoes damped oscillations.

Since the structure formation occurs in the decelerating phase,
we can address the question within Newtonian theory
by generalizing the spherical model.
In the case of vanishing shear and rotation,
the continuity and Euler-Poisson equations become
\begin{equation}
\dot{\rho} + 3 {\cal{H}} \, \rho  =  0\, ;  \;\;\;\;\;\;
3 \dot{\cal{H}} + 3 {\cal{H}}^{2} +
4 \pi G \rho + \vec{\nabla} \cdot
\left( \frac{v^{2}}{\rho}\, \vec{\nabla}  \rho \right)  =  0 ,
\label{eq007b}
\end{equation}
where $\cal{H}$ is the local Hubble parameter.
It is reasonable to approximate
$\delta (t,\vec{x}) \equiv \rho (t,\vec{x})/\bar{\rho}(t)-1$
   by  the spherical lump
    $\delta (t,\vec{x}) = \delta_{k}(t) \sin (kx)/
    (kx)$ .
We then find that $\delta_{k} (a)$ satisfies~\cite{bil8}
\begin{equation}
a^{2}  \delta_{k}^{''} + \frac{3}{2}\, a\, \delta_{k}^{'} -
\frac{3}{2}\,\delta_{k} (1 + \delta_{k})-\frac{4}{3}\,
\frac{(a \,\delta_{k}^{'})^{2}}{1 + \delta_{k}} + \frac{49}{4} \,
 \left( \frac{a}{a_{k}} \right)^{7}
\frac{\delta_{k}}{(1 + \delta_{k})^{2}}  = 0  .
\label{eq008}
\end{equation}
where
$a_k= (d_{\rm{s}}k )^{-2/7} a$.
Eq.\ (\ref{eq008})
reproduces (\ref{eq004}) at linear order and
extends the spherical dust model by
incorporating the Jeans length through the last term.

\begin{figure}
\begin{minipage}[t]{0.4\linewidth}
\begin{center}
\includegraphics[width=0.87\textwidth,trim= 0 2.3cm 0 2cm]{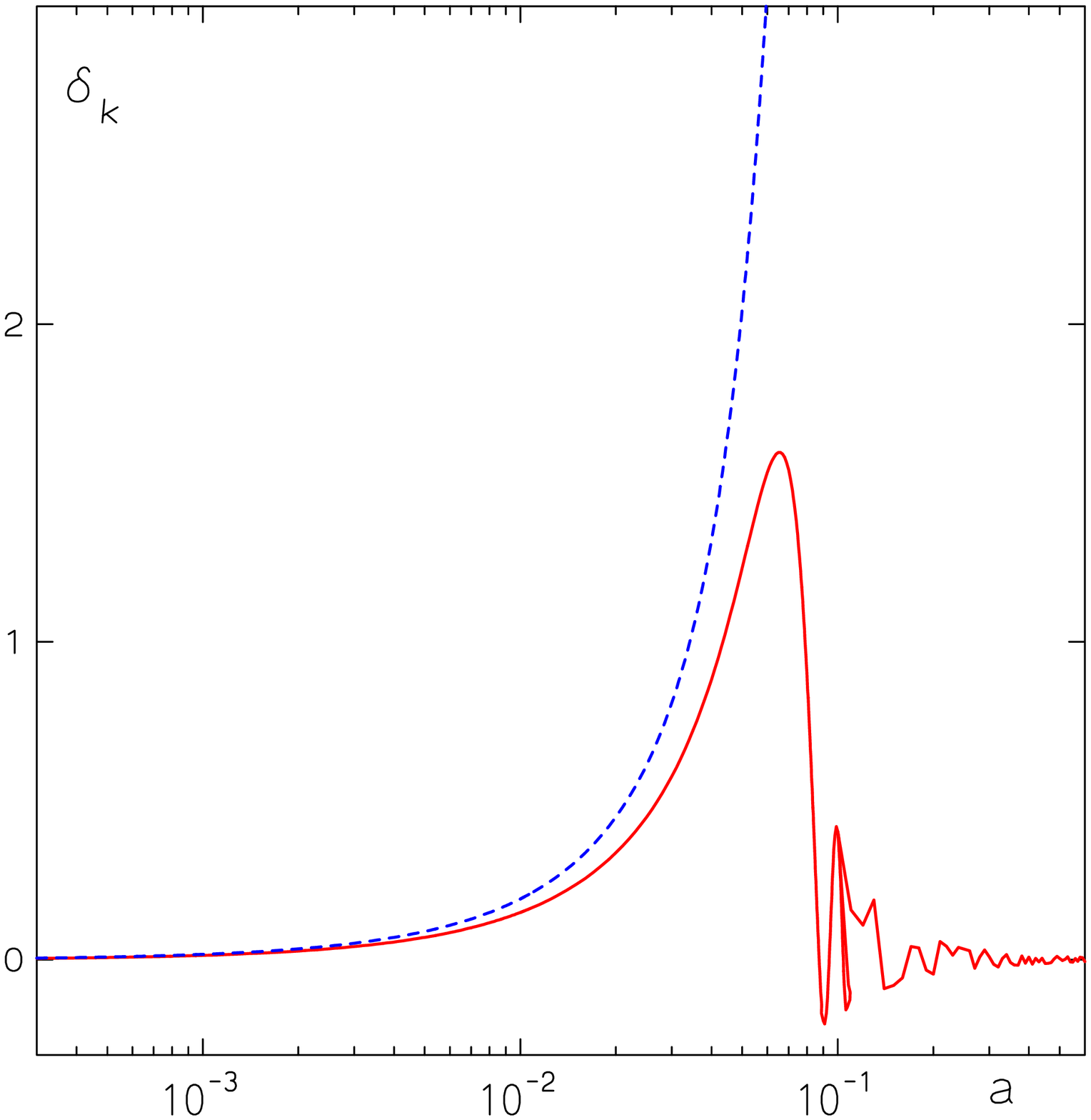}
\end{center}
\end{minipage}
\begin{minipage}[t]{0.6\linewidth}
\begin{center}
\includegraphics[width=0.9\textwidth,trim=0 0 0 2cm]{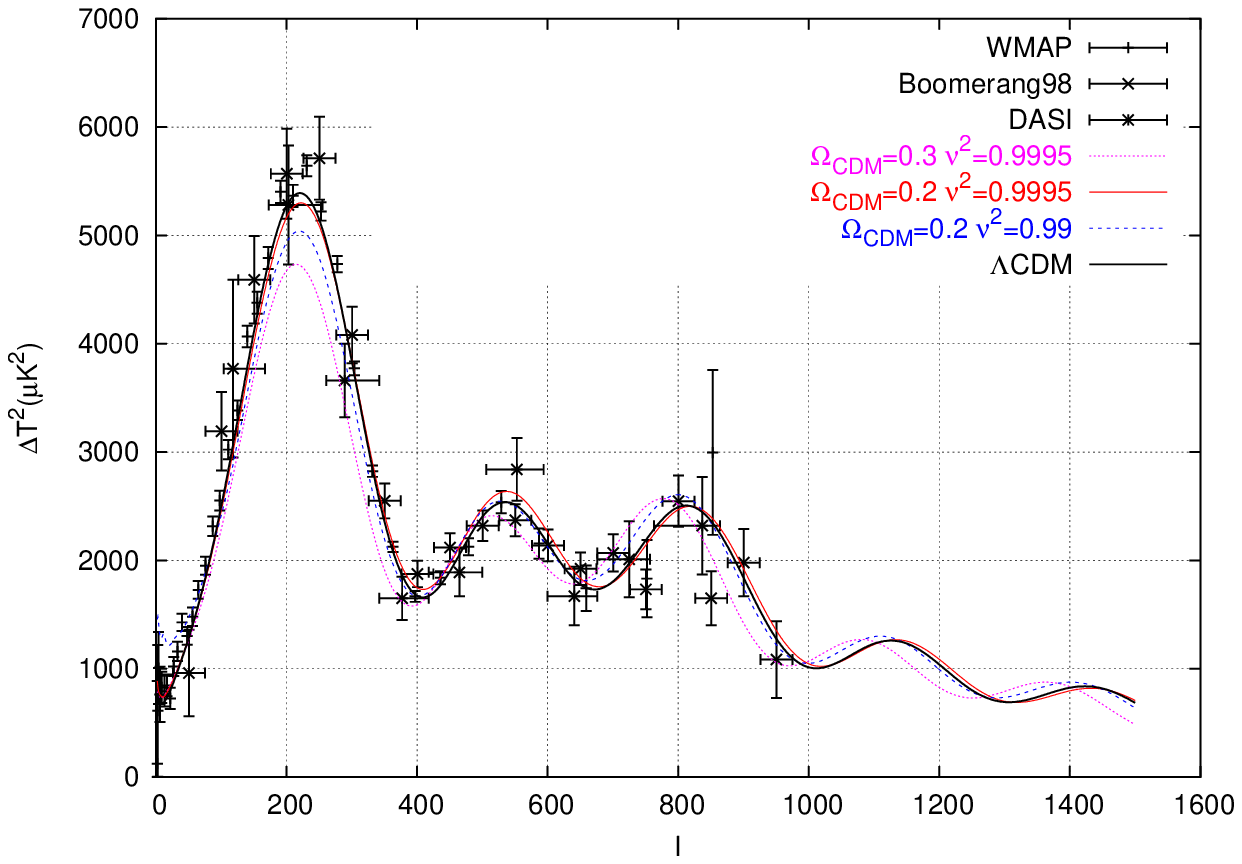}
\end{center}
\end{minipage}
\begin{minipage}[t]{0.4\linewidth}
\begin{center}
\includegraphics[width=0.88\textwidth,trim= 0 2.3cm 0 2cm]{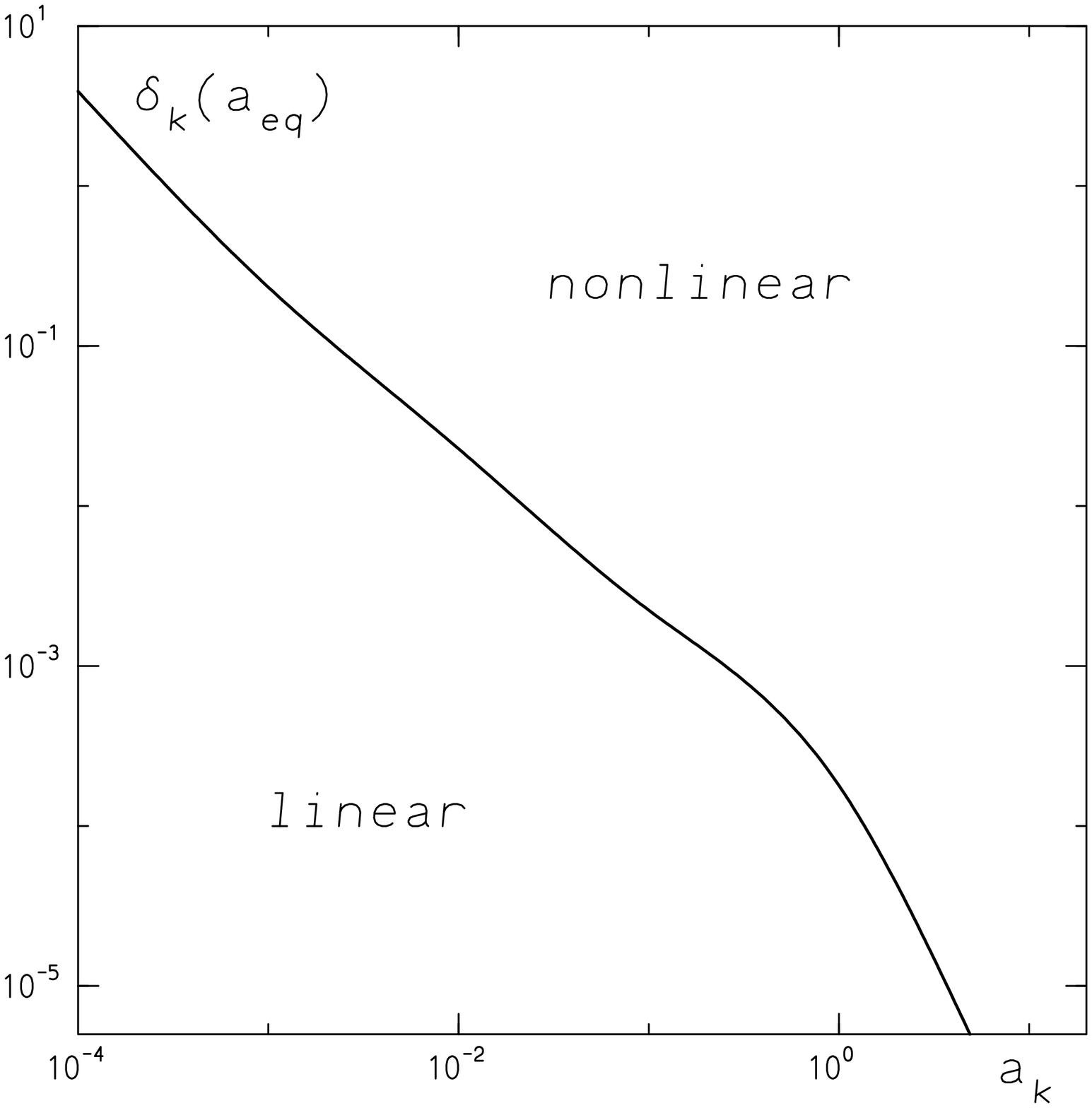}
\end{center}
\end{minipage}
\begin{minipage}[t]{0.6\linewidth}
\begin{center}
\includegraphics[width=0.9\textwidth,trim=0 0 0 2cm]{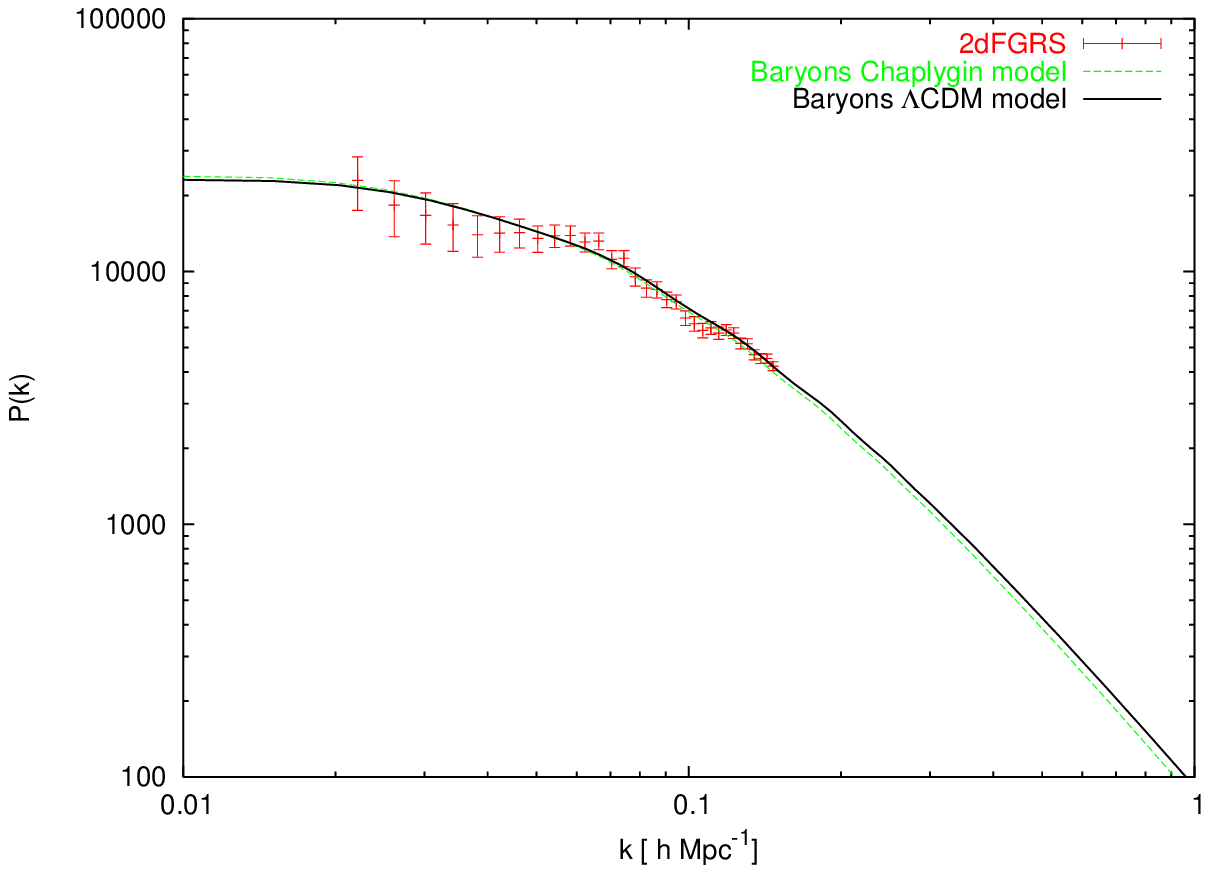}
\end{center}
\end{minipage}
\caption{(a) Top left:
 Evolution of $\delta_{k}(a)$ in
            the spherical model, Eq.\ (\ref{eq008}),
           from $a_{\rm eq} = 3\times 10^{-4}$
           for $a_{k}$ = 0.0476,
           $\delta_{k} (a_{\rm eq})$ =0.004 (solid)
            and $\delta_{k} (a_{\rm eq})$ =0.005 (dashed).
(b) Bottom left:
The critical $\delta_k(a_{\rm eq})$ versus $a_k$.
(c) Top right:
 CMB angular power spectrum for a mixture of Chaplygin gas and
 condensate with $\Omega_{\rm gas}$ = 0.7 and $h$ = 0.7,
varying $\Omega_{\rm CDM}$ and $v^{2}$.
(c) Bottom right:
Baryon power
spectra for the mixed Chaplygin gas
 with
 $\Omega_{\rm CDM}=0.2$ and $v^{2}=0.9995$
 and for
 the $\Lambda$CDM model.
A normalization factor of 1.35 is used.
}\label{fig1}
\end{figure}

In Fig.\ \ref{fig1}a we show the evolution of two initial
perturbations from radiation-matter equality
for $a_{k}=a_{\rm reion} = 1/21$.
One sees that for sufficiently small
$\delta_{k} (a_{\rm eq})$, the acoustic horizon can stop
$\delta_{k} (a)$ from growing even in a mildly nonlinear regime.
Conversely, for $\delta_{k} (a_{\rm eq})$ above the
threshold, $\delta_{k} (a) \rightarrow \infty$ at
finite $a$ just as in the dust model.
The critical $\delta_k(a_{\rm eq})$
dividing the two regimes depends strongly on
$a_k$
(Fig.\ \ref{fig1}b).
Qualitatively similar conclusions have been reached
in a different way
by Avelino {\it et al}
\cite{ave}.
Since the critical $\delta_k$ is commensurate with
the peak in the conditional probability distribution for
spheroidal collapse \cite{lee24}, and $a_{k}$
is only weakly dependent on the comoving
wave number, we can thus be confident
that the Chaplygin gas will evolve at high redshift
into a mixed system
consisting of smoothly
distributed gas and gravitational condensate.
The latter will participate
in hierarchical clustering as CDM.

Homogeneous world models containing a mixture
of cold dark matter and Chaplygin gas
have been successfully confronted with
lensing statistics \cite{dev15} as well as
with supernova and other tests \cite{avel16}.
Clearly, it is only a matter of interpretation to replace
`cold dark matter' by `Chaplygin droplet matter' (condensate).
On the other hand, ``unification'' has often been misconstrued
to mean a mixture of baryonic and perturbative Chaplygin gas
only \cite{avel16}$^-$\cite{sand18}.
Owing to the damped oscillation of perturbations and the driving decay
of the gravitational potential,
it is hardly surprising that the baryon
plus the purely perturbative Chaplygin gas model
is in gross conflict with the CMB~\cite{bent2,cat17} and
mass power spectrum \cite{sand18} data.
Hence we have undertaken a calculation of the
CMB anisotropies and the mass power spectrum
in our unification scenario based on
the Chaplygin gas and including  nonlinear condensate.
As the sonic horizon is
negligible at recombination and the droplets
affect the CMB and the power spectrum
only through the gravitational potential, it
is adequate to treat them throughout as ordinary cold dark matter,
parametrized by $\Omega_{\rm CDM}$, in the perturbation equations.
The residual uncondensed gas
is parametrized as
\begin{equation}
\bar{\rho}_{\rm gas} (a) = \rho_{\rm cr} \Omega_{\rm gas} \; 
\left[ v^{2} + (1 - v^{2})/a^{6} \right]^{1/2}
\label{eq009}
\end{equation}
in the background and
\vspace{-0.3cm}
\begin{equation}
\bar{v}^{2} (a) = v^{2}\left[v^{2} +
 (1 - v^{2})/ a^{6} \right]^{-1}
\label{eq010}
\end{equation}
in the perturbation equations.
The parameters $\Omega_{\rm gas}$ and $v$
     are not unrelated.
From the statistical
 distribution of the eigenvalues of the
deformation tensor it follows that
only in 8\% of the cases
there is expansion along all three principal axes \cite{zel13}, hence
only about 8\% of the initial Chaplygin gas should fail to condense.
Therefore, we expect the initial fraction of  uncondensed gas
      to be
\begin{equation}
\Omega_{\rm gas} \, \sqrt{1 - v^{2}}\,(\Omega_{\rm CDM}
+ \Omega_{\rm gas} \, \sqrt{1 - v^{2}})^{-1}\simeq 0.08 .
\label{eq011}
\end{equation}
This equation gives an estimate of the parameter $v^2$ in terms of
$\Omega_{\rm gas}$.

In Fig.\ \ref{fig1}c we compare the
CMB angular power spectrum obtained by implementing
(\ref{eq009}) and (\ref{eq010})
in a modification of the CMBfast \cite{sel20} code
with the WMAP data \cite{ben4}.
Albeit the result is preliminary,
and by no means a best fit, it is apparent that
the CMB data can be described by an evolved mixture of
Chaplygin gas and condensate
with parameters satisfying Eq.\ (\ref{eq011}).

In Fig.\ \ref{fig1}d we exhibit the baryon
power spectrum calculated in the two models:
the mixed Chaplygin gas
and the $\Lambda$CDM
with the optimal CMB parameters of
Fig.\  \ref{fig1}c.
The model spectra have been convolved with
the 2dFGRS
window function and their amplitude fitted
to the power spectrum data
\cite{2df}.
Again, the data is well described in the range of $k$
in which the window function is available.
Hence, it may safely be said
that unification stands up as a viable scenario.


\section*{References}
\begin{thebibliography}{99}
\bibitem{jack8} R.\ Jackiw,
{\em Lectures on fluid dynamics}
(Springer-Verlag, New York, 2002).
\bibitem{kam9} A.\ Kamenshchik, U.\ Moschella and V.\ Pasquier,
{\em Phys.\ Lett.} B
{\bf 511}, 265 (2001).
\bibitem{bil6} N.\ Bili\'{c}, G.B.\ Tupper and R.D.\ Viollier,
{\em Phys.\ Lett.} B {\bf 535}, 17 (2002).
\bibitem{bil7} N.\ Bili\'{c}, G.B.\ Tupper and R.D.\ Viollier,
 astro-ph/0207423.
\bibitem{sen} A.\ Sen, {\em Mod.\ Phys.\ Lett.} A {\bf 17},
1797 (2002);
{\em JHEP} {\bf 0204}, 048 (2002).
\bibitem{zel13} Ya.B.\ Zel'dovich,
{\em Astron.\ Astrophys.} {\bf 5}, 84 (1970).
\bibitem{fab14} J.C.\ Fabris, S.V.B.\ Goncalves and P.E.\ de Souza,
{\em Gen.\ Rel.\ Grav.} {\bf 34}, 53 (2002).
\bibitem{bil8}
N.\ Bili\'{c}, R.J.\ Lindebaum, G.B.\ Tupper and R.D.\ Viollier,
astro-ph/0307214.
\bibitem{ave} P.P.\ Avelino {\it et al},
astro-ph/0306493.
\bibitem{lee24} J.\ Lee and S.F.\ Shandarin,
{\em Astrophys.\ J.} {\bf 500}, 14 (1998).
\bibitem{dev15} A.\ Dev, J.S.\ Alcaniz and D.\ Jain,
{\em Phys.\ Rev.} D {\bf 67}, 023515 (2003).
\bibitem{avel16} P.P.\ Avelino {\it et al},
{\em Phys.\ Rev.} D {\bf 67}, 023511 (2003).
\bibitem{bent2} M.C.\ Bento, O.\ Bertolami, and A.A.\ Sen,
{\em Phys.\ Rev.} D {\bf 67}, 063003 (2003);
\bibitem{cat17} P.\ Carturan and F.\ Finelli,
{\em JCAP} {\bf 0307}, 005 (2003),
astro-ph/0211626;
L.\ Amendola,  F.\ Finelli, C.\ Burigana, and D.\ Carturan,
astro-ph/0304325.
\bibitem{sand18} H.B.\ Sandvik, M.\ Tegmark,
M.\ Zaldarriaga and I.\ Waga, astro-ph/0212114.
\bibitem{sel20} U.\ Seljak and M.\ Zaldarriaga,
{\em Astrophys.\ J.} {\bf 469}, 437 (1996).
\bibitem{ben4} C.L.\ Bennett {\it et al}, astro-ph/0302207;
D.N.\ Spergel {\it et al}, astro-ph/0302209.
\bibitem{2df} W.J.\ Percival {\it et al},
{\em MNRAS} {\bf 327}, 1297 (2001).
\end{thebibliography}
\end{document}